\def\jms{J. Mol.\ Spectrosc.\ }

\def\mnras{Mon.\ Not.\ Roy.\ Astron.\ Soc.\ }
\def\cjp{Can.\ J. Phys.\ }
\def\aa{Astron.\ Astrophys.\  }
\def\aass{Astron.\ Astrophys.\ Suppl.\ Ser.\ }

\documentclass[aps,prl,preprint,superscriptaddress,showpacs]{revtex4}
\usepackage{graphics}

\begin{document}

\title{
First constraint on cosmological variation of the proton-to-electron mass ratio from two independent telescopes}

\author{F. van Weerdenburg}
\affiliation{Institute for Lasers, Life and Biophotonics, VU University Amsterdam, de Boelelaan 1081, 1081 HV Amsterdam, The Netherlands}
\affiliation{Astronomical Institute Anton Pannekoek, Universiteit van Amsterdam,
1098 SJ Amsterdam, The Netherlands}
\author{M. T. Murphy}
\affiliation{Centre for Astrophysics and Supercomputing,
Swinburne University of Technology, Melbourne, Victoria 3122, Australia}
\author{A. L. Malec}
\affiliation{Centre for Astrophysics and Supercomputing,
Swinburne University of Technology, Melbourne, Victoria 3122, Australia}
\author{L. Kaper}
\affiliation{Institute for Lasers, Life and Biophotonics, VU University Amsterdam, de Boelelaan 1081, 1081 HV Amsterdam, The Netherlands}
\affiliation{Astronomical Institute Anton Pannekoek, Universiteit van Amsterdam,
1098 SJ Amsterdam, The Netherlands}
\author{W. Ubachs}
\affiliation{Institute for Lasers, Life and Biophotonics, VU University Amsterdam, de Boelelaan 1081, 1081 HV Amsterdam, The Netherlands}
\date{\today}

\begin{abstract}

A high signal-to-noise spectrum covering the largest number of hydrogen lines (90 H$_2$ lines and 6 HD lines) in a high redshift object was analyzed from an observation along the sight-line to the bright quasar source J2123$-$005 with the UVES spectrograph on the ESO Very Large Telescope (Paranal, Chile). This delivers a constraint on a possible variation of the proton-to-electron mass ratio of $\Delta\mu/\mu = (8.5 \pm 3.6_{\mathrm{stat}} \pm 2.2_{\mathrm{syst}})\times 10^{-6}$ at redshift $z_{\mathrm{abs}}=2.059$, which agrees well with a recently published result on the same system observed at the Keck telescope yielding $\Delta\mu/\mu = (5.6 \pm 5.5_{\mathrm{stat}} \pm 2.9_{\mathrm{syst}})\times 10^{-6}$. Both analyses used the same robust absorption line fitting procedures with detailed consideration of systematic errors.

\end{abstract}


\maketitle

Experimental searches for temporal variation of fundamental constants on a cosmological time scale
have resurfaced on the agenda of contemporary physics since the ground-breaking study of a possible
variation of the fine-structure constant $\alpha$ from quasar observations, now a decade ago~\cite{Webb1999}. While $\alpha=e^2/4\pi\epsilon_0\hbar c$ is a fundamental constant representing the strength of the electromagnetic force, the proton-to-electron mass ratio $\mu=m_p/m_e$ is another important dimensionless constant to be tested for possible temporal drifts. Since the quark masses contribute only marginally to the proton mass ($\sim 10$\%), $m_p$ is nearly proportional to the strength of the nuclear force, $\Lambda_{QCD}$. Hence, $\mu$ scales with $\Lambda_{QCD}/v_H$, $v_H$ being the Higgs vacuum expectation value. Therefore, $\Delta\mu/\mu$ probes the cosmological evolution of the nuclear vs. the electroweak sector in the Standard Model. Interestingly, various models of Grand Unification predict relations of the form
\begin{equation}
\Delta\mu/\mu = R_{\mu\alpha} \Delta\alpha/\alpha
\label{Eq-R}
\end{equation}
where the proportionality constant is normally large, \emph{i.e.} $|R_{\mu\alpha}|\,\approx 10-40$~\cite{Flambaum2004}, although model dependent~\cite{Dent2008}. Measuring any possible changes in $\alpha$ and $\mu$ therefore allows a test of `beyond Standard' theories.

At cosmological redshifts $z>2$ molecular hydrogen is the test system for probing a drift in $\mu$~\cite{Thompson1975,Varshalovich1993,Ubachs2007}, where the time arrow is defined by $\Delta\mu=\mu_z - \mu_0$. Wavelengths $\lambda^z$ of spectral lines in the Lyman and Werner bands of H$_2$ observed in distant objects, may be compared to the wavelengths $\lambda^0$ for the same lines measured in the laboratory at high accuracy~\cite{Ubachs2007,Reinhold2006,Salumbides2008} via
\begin{equation}
\frac{\lambda_i^z}{\lambda_i^0}=(1+z_{\mathrm{abs}}) \left(1 + K_i \frac{\Delta\mu}{\mu}\right)
\label{Eq-K}
\end{equation}
with $z_{\mathrm{abs}}$ the redshift of the absorbing gas cloud in the line-of-sight of the quasar and $K_i$ the so-called sensitivity coefficient, expressing the sensitivity for each spectral line to a mass variation $\Delta\mu$; the $K_i$ can be calculated to an accuracy of 1\%~\cite{Thompson1975,Reinhold2006,Ubachs2007}. For the deuterated isotopologue HD, accurate laboratory data and $K$-coefficients are available~\cite{Ivanov2008}.

\begin{figure*}
\resizebox{1.0\textwidth}{!}{\includegraphics{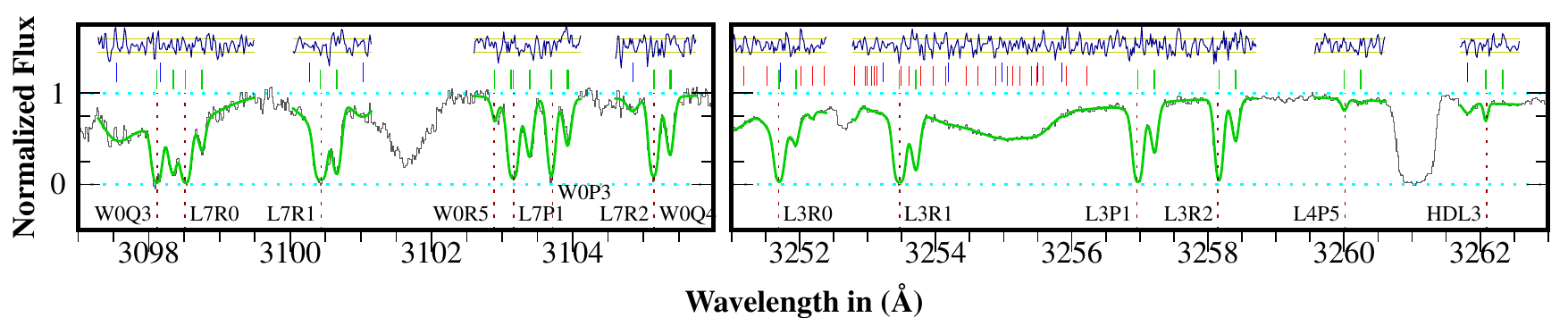}}
\caption{(Color online) Spectrum of J2123$-$005 observed with UVES-VLT clearly showing several H$_2$ transitions, each comprising two distinct velocity features. Left: region $3097-3106$ \AA\ with relatively low SNR. Right: region $3251-3263$ \AA\ toward longer wavelength at higher SNR, also containing a line of HD. The upper trace shows residuals in the fit regions. The tick marks above the spectrum refer to the velocity components of the H$_2$ and HD lines (green), modeled Lyman-$\alpha$ forest lines (blue), and Fe {\sc ii} lines (red). H$_2$ spectral line identifications are shown at the bottom.}\label{Fig1}
\end{figure*}

Here we report on an investigation of the H$_2$ absorption system at $z_{\mathrm{abs}}=2.06$ toward the quasar J2123$-$005 making use of the Ultraviolet and Visible Echelle Spectrograph (UVES) at the Very Large Telescope (ESO, Paranal, Chile). The same object was recently investigated with the HIRES instrument at the Keck Telescope~\cite{Malec2010}, so that now, for the first time, an accurate $\mu$-variation analysis becomes available from two independent telescopes. Moreover, J2123$-$005 is the brightest quasar with H$_2$ absorption, with $R=15.8$ mag. An 11h observation in August and September 2008 at VLT delivered a spectrum reaching a signal-to-noise ratio of $~65$ per $1.5$ km s$^{-1}$ pixel at 3415 \AA, degrading to $\sim 15$ at the far UV end (3046 \AA). In this range many hydrogen lines are found, of which 90 H$_2$ and 6 HD are useful to include in an analysis; two typical spectral regions are shown in Fig.~\ref{Fig1}, while the entire spectrum is reproduced in the additional material~\cite{Additional}. Individual spectra ($14$ exposures of $2900$s) were air-vacuum and heliocentrically corrected, rebinned and coadded.

This high-quality VLT spectrum of J2123 was subjected to a comprehensive fitting procedure with a direct $\chi^2$ minimization of $\Delta\mu/\mu$ to the entire spectrum, in a similar manner as in~\cite{King2008,Malec2010}. This procedure has many advantages over a line-by-line fitting procedure as used in other studies~\cite{Reinhold2006,Ivanchik2005,Thompson2009}. The former allows more objective selection of spectral regions to be fitted and lines that are partly blended by Lyman-$\alpha$ forest lines can be modeled and explicitly included in the fitting routine. By this means the amount of information on H$_2$ absorption extracted from the spectrum is effectively increased; however, the regions where no H$_2$ absorption occurs are cut out of the fitting routine to avoid contaminating the $\chi^2$. In total, $68$ regions cover the $96$ H$_2$/HD lines included in the fit. Additional regions containing some $39$ H$_2$/HD lines were excluded, because lines were too weak, or located within a saturated (Lyman-$\alpha$) absorption region, or were contaminated with unidentified narrow spectral features, \emph{i.e} metal lines.

A generic model spectrum is produced from the accurately known transition wavelengths for H$_2$~\cite{Salumbides2008} and HD lines~\cite{Ivanov2008}. Intensities derive from multiplying calculated oscillator strengths~\cite{Abgrall1994} with column densities $N(J)$ for each rotational state as fitting parameters. This mask spectrum is then convolved threefold, with a Lorentzian for the damping coefficient, $\Gamma_i$, for each line~\cite{Abgrall2000}, a fixed width parameter for the instrumental profile of the UVES spectrograph ($0.8"$ resulting in a resolving power of 53000), and a Gaussian fit parameter, $b$, for the Doppler width. In this way the H$_2$ and HD lines in the spectrum are tied, leading to a reduction in the number of parameters to be adjusted in the fitting procedure.

This spectrum is replicated to accommodate the velocity structure of the absorbing cloud, starting with the two features split by $\sim 25$ km/s visible in Fig.~\ref{Fig1}. Subsequently, additional velocity components (VCs) are added, thereby increasing the number of fit parameters by $b_\mathrm{vc}$ and $N_\mathrm{vc}(J)$ for all VCs. Only after a full H$_2$/HD velocity structure is imposed, all Lyman-forest and metal lines are assigned and simulated properly, and the continuum and zero levels are adjusted for all local fitting regions, a possible variation of the proton-to-electron mass ratio $\Delta\mu/\mu$ is included as a final fitting parameter. The VPFIT algorithm~\cite{VPFIT} uses Eq.~(\ref{Eq-K}) and the $K_i$ coefficients~\cite{Ubachs2007,Reinhold2006,Malec2010} directly. In addition to Lyman-$\alpha$ forest lines, metal lines falling in the H$_2$ regions are also co-fitted. As one example we mention the 18 velocity components of the Fe~{\sc ii} line, weakly overlapping the H$_2$ L3R0 line at $3252$ \AA~\cite{Additional}; its velocity structure is tied in the fit to the Fe~{\sc ii} line observed at $7952$ \AA\ ($\lambda_{\mathrm{lab}}=2600$ \AA).

\begin{figure}[b]
\resizebox{0.47\textwidth}{!}{\includegraphics{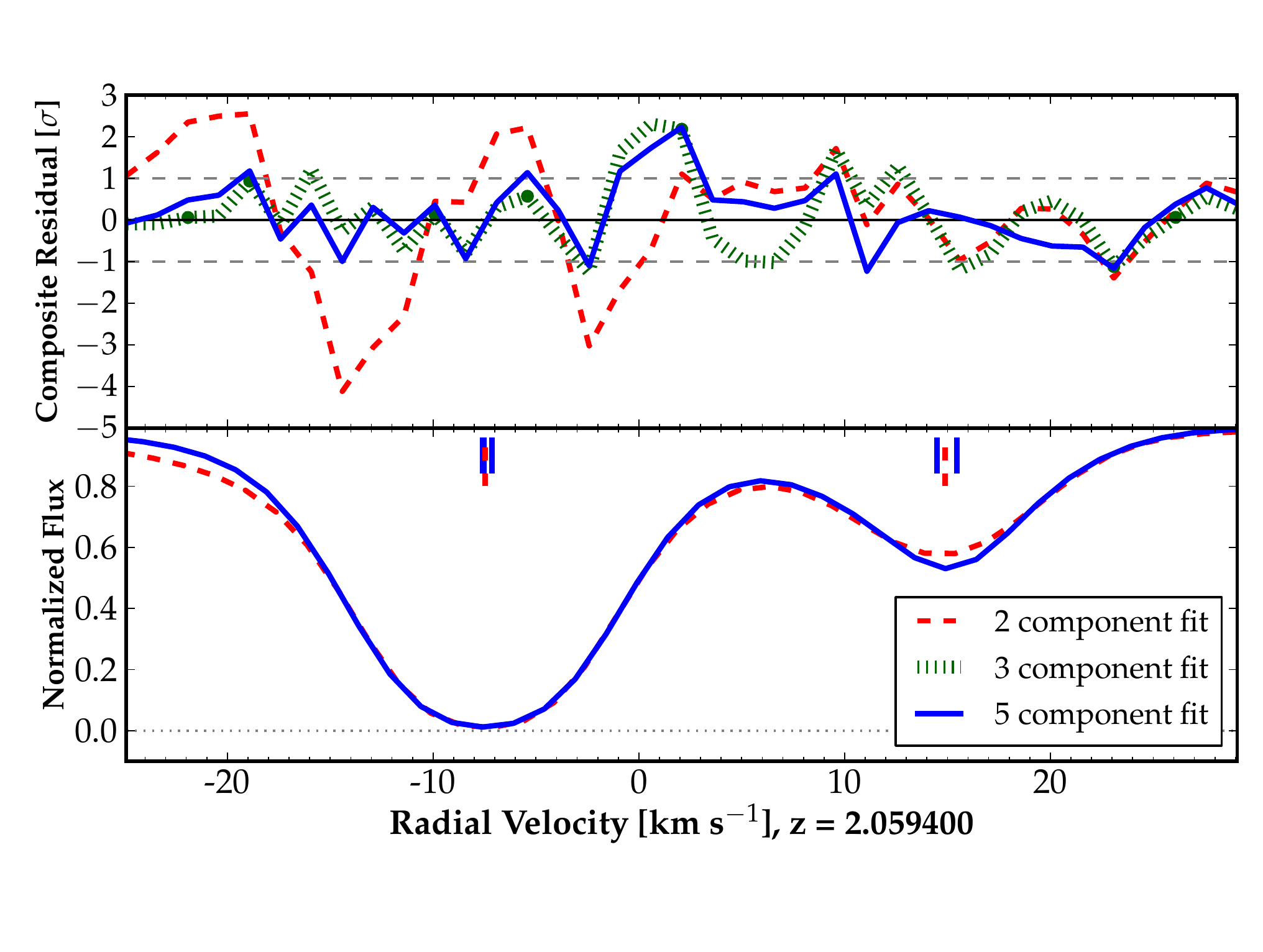}}
\caption{(Color online) Composite residual spectrum composed of the 25 cleanest H$_2$ lines for fits to 2, 3 and 5 velocity components. Upper panel: residuals. Lower panel: the spectrum as simulated from the final parameters resulting in the 2 and 5 VC models. The tick marks indicate the fitted positions in 2 and 5 VC models; in the 5 VC model two VCs lie at almost the same velocity and appear as a single tick mark on this plot.}\label{Fig2}
\end{figure}

\begin{figure}
\resizebox{0.47\textwidth}{!}{\includegraphics{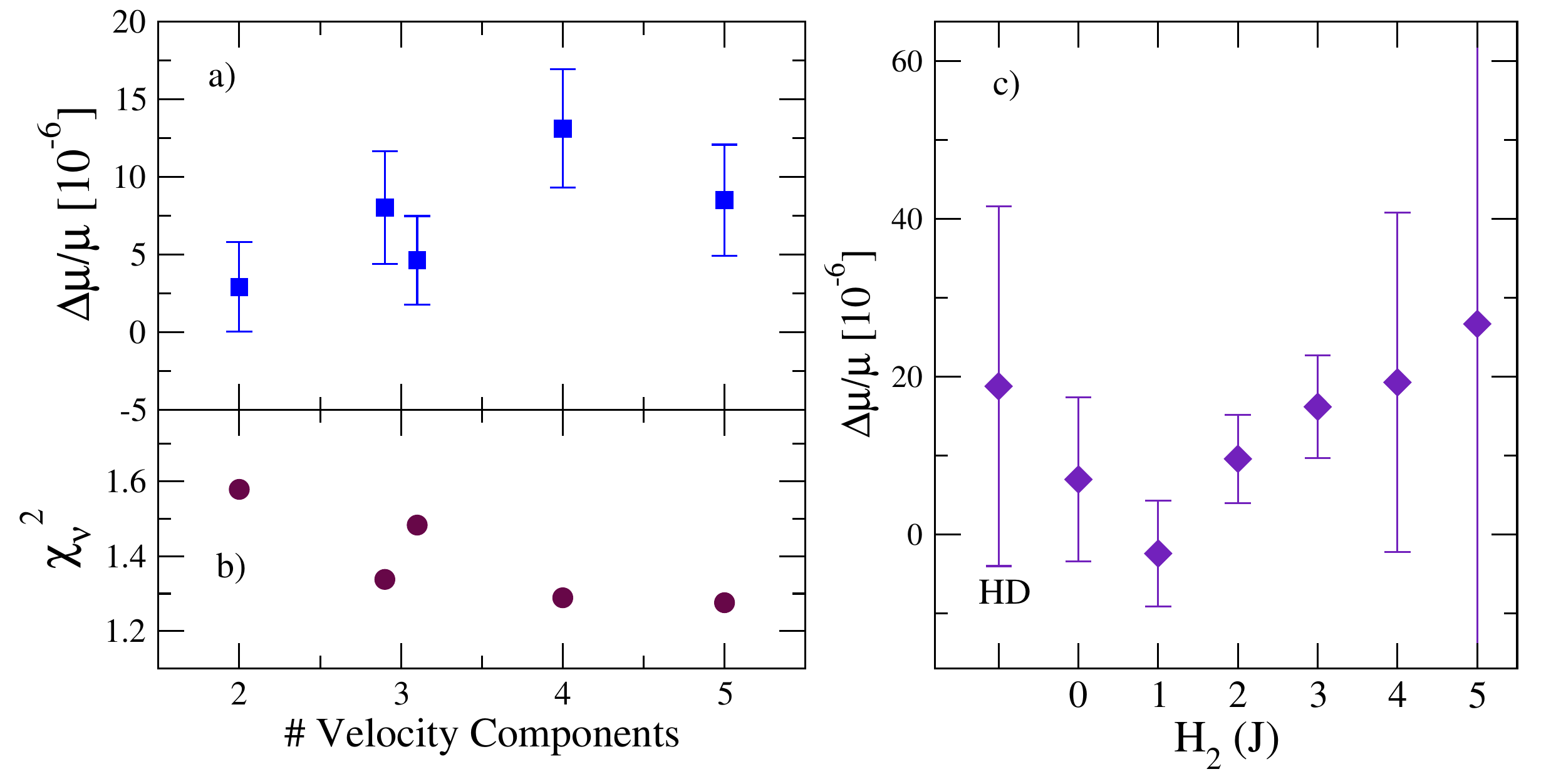}}
\caption{(Color online) Results from various fits. a) resulting value for $\Delta\mu/\mu$ for progressively increasing the number of velocity components; there are two results for 3VC fits, obtained by putting the 3$^{\rm{rd}}$ component in both distinct velocity features, which are horizontally offset for clarity; b) the $\chi_{\nu}^2$ value; c) result on $\Delta\mu/\mu$ from fits to 7 subclasses of spectral lines. }\label{Fig3}
\end{figure}

One of the advantages of fitting all lines simultaneously is that the velocity structure of the absorbing system can be robustly investigated. For example, a composite residual spectrum was constructed from the 25 cleanest absorption lines, all shifted to a common redshift ($z=2.059400$), as shown in Fig.~\ref{Fig2}. It demonstrates that significant H$_2$ absorption is left in a systematic way in the two-component fit; four and a five-component fits clearly yield better representations of the spectrum. Besides visual inspection of residuals, the procedure of adding velocity components can be quantified by calculating the value of $\chi^2$ per degree of freedom, $\chi^2_{\nu}$, as is shown in Fig.~\ref{Fig3}. $\chi^2_{\nu}$ decreases until a possible $6^{\mathrm{th}}$ VC, which is then rejected on the basis of statistical grounds. Application of a modified version of the Akaike Information Criterion for finite sample sizes, yields a value of $\Delta_{AIC} \sim 90$ (for definition and evaluation see~\cite{Sigiura1978}), thus establishing an objective ground for adding a $5^{\mathrm{th}}$ VC, and for leaving out the $6^{\mathrm{th}}$ VC. The result of the $5$-component fit is thus obtained as the fiducial statistical outcome with $\Delta\mu/\mu = (8.5 \pm 3.6)\times 10^{-6}$.

The robustness of the statistical analysis was further investigated by verifying that the result was not produced by a limited subclass of lines. A fit was performed whereby a different value of $\Delta\mu/\mu$ was fitted simultaneously to each of the rotational states in hydrogen, H$_2$($J$), and for HD. The results shown in Fig.~\ref{Fig3}c demonstrate the consistency for the value of $\Delta\mu/\mu$ in this procedure. In the model description the physical assumption was made that the Doppler widths, $b_\mathrm{vc}$ (for each velocity component $z_\mathrm{vc}$), were independent of the rotational state; if, however, a fit is performed with varying Doppler widths $b_v(J)$, a $\Delta\mu/\mu$ value is returned differing by only 5\%.

The analysis strategy aimed to stick as close as possible to the known molecular physics of H$_2$ and HD. The damping parameters, $\Gamma_i$, from \cite{Abgrall2000} describe the wings of the saturated lines (as in Fig.~\ref{Fig1}) well. The oscillator strengths, $f_i$, were kept fixed to the molecular physics values~\cite{Abgrall1994}, thereby keeping the number of fit parameters as low as possible. However, in the region $3345 - 3415$ \AA, covering the L0 and L1 Lyman lines, the calculated $f_i$ underestimate the line intensities by some 50\%. Hence, in the fits the oscillator strengths in this range were adjusted, similarly as in~\cite{Malec2010}, where this phenomenon was attributed to the O~{\sc iv}/Lyman-$\beta$ emission feature of the background quasar.

As for systematic effects on the resulting value of $\Delta\mu/\mu$, adverse influences on the wavelength calibration are of particular importance. Wavelength calibration of the composite spectrum was derived from Th-Ar frames attached to each science exposure, where use was made of the improved Th-Ar line selection algorithm rejecting blended and weak lines~\cite{Murphy2007}. Wavelength calibration residuals of Th-Ar lines, used for calibrating the echelle grating are typically $70$ m/s. However, since the wavelength scale of each echelle order is derived from typically more than 11 Th-Ar lines and many H$_2$ transitions are fitted for each order, this effect largely averages out, and long-range variations of the wavelength scale are limited to 30 m/s~\cite{Murphy2007}. In view of a span in $K_i$ of $\sim 0.05$, a shift of $30$ m/s translates into a maximum variation in $\Delta\mu/\mu$ of $2.0 \times 10^{-6}$; here velocity shifts are expressed as $\Delta v = cK_i \Delta\mu/\mu$.

In addition, intra-order distortions within individual echelle orders, first studied for the HIRES-Keck
grating system~\cite{Griest2010} and later for UVES-VLT~\cite{Whitmore2010}, may result
in peak-to-peak velocity shifts of up to $\Delta v=200$ m/s between transitions. This phenomenon was simulated by producing a counter-distorted spectrum by shifting the center of all echelle orders by $-100$ m/s and imposing shifts up to $+100$ m/s at the order edges. From a fit to this spectrum it is estimated that this phenomenon produces a possible systematic effect of $0.7 \times 10^{-6}$. Systematic effects from spectral re-dispersion occurring in the re-binning process when averaging over the 14 science exposures are less than $0.2 \times10^{-6}$. Finally, drifts in temperature and atmospheric pressure between the quasar and Th-Ar calibration exposures are $< 1$ K and $< 1$ mbar, respectively, leading to a possible error of  $< 0.7 \times 10^{-6}$. Taking the possible shifts in quadrature this results in an estimate of the systematic error of $\delta_{\mathrm{syst}}(\Delta\mu/\mu)= 2.2 \times 10^{-6}$. Uncertainties in the $K_i$ coefficients may lead to uncertainties in $\Delta\mu/\mu$ at the $\sim 0.01 \times\Delta\mu/\mu$ level~\cite{Ubachs2007}, \emph{i.e.} they are negligible.

A $\Delta\mu$-effect can be mimicked by a longe-range distortion of the wavelength scale due to the correlation of $K$-coefficients with wavelength~\cite{Ubachs2007}. Separate fitting of the range $\lambda < 3230$ \AA, where Werner and Lyman lines exhibiting strongly differing $K_i$ values are found alongside each other, returns a value within $10$\% of the fiducial value, although with a larger uncertainty margin. This can be considered as an internal consistency check of the wavelength scale.

The statistical and systematic analyses lead us to a final result for the UVES-VLT H$_2$ absorption spectrum toward J2123 (uncertainties representing $1\sigma$):
\begin{equation}
 \Delta\mu/\mu = (8.5 \pm 3.6_{\mathrm{stat}} \pm 2.2_{\mathrm{syst}})\times 10^{-6}.
\end{equation}
This may be compared to the result from the analysis of the HIRES-Keck spectrum for the same object~\cite{Malec2010}:
\begin{equation}
 \Delta\mu/\mu = (5.6 \pm 5.5_{\mathrm{stat}} \pm 2.9_{\mathrm{syst}})\times 10^{-6}.
\end{equation}
These values are well within $1\sigma$ of each other and may be averaged to yield $\Delta\mu/\mu = (7.6 \pm 3.5_{\mathrm{tot}})\times 10^{-6}$.  The agreement obtained between both investigations of the same high redshift H$_2$ absorption spectrum from the two major large telescopes VLT and Keck enhances the confidence in the statistical and systematic uncertainty analyses. They are indicative of the error sources being suitably taken into account for both instruments.

\begin{figure}
\resizebox{0.47\textwidth}{!}{\includegraphics{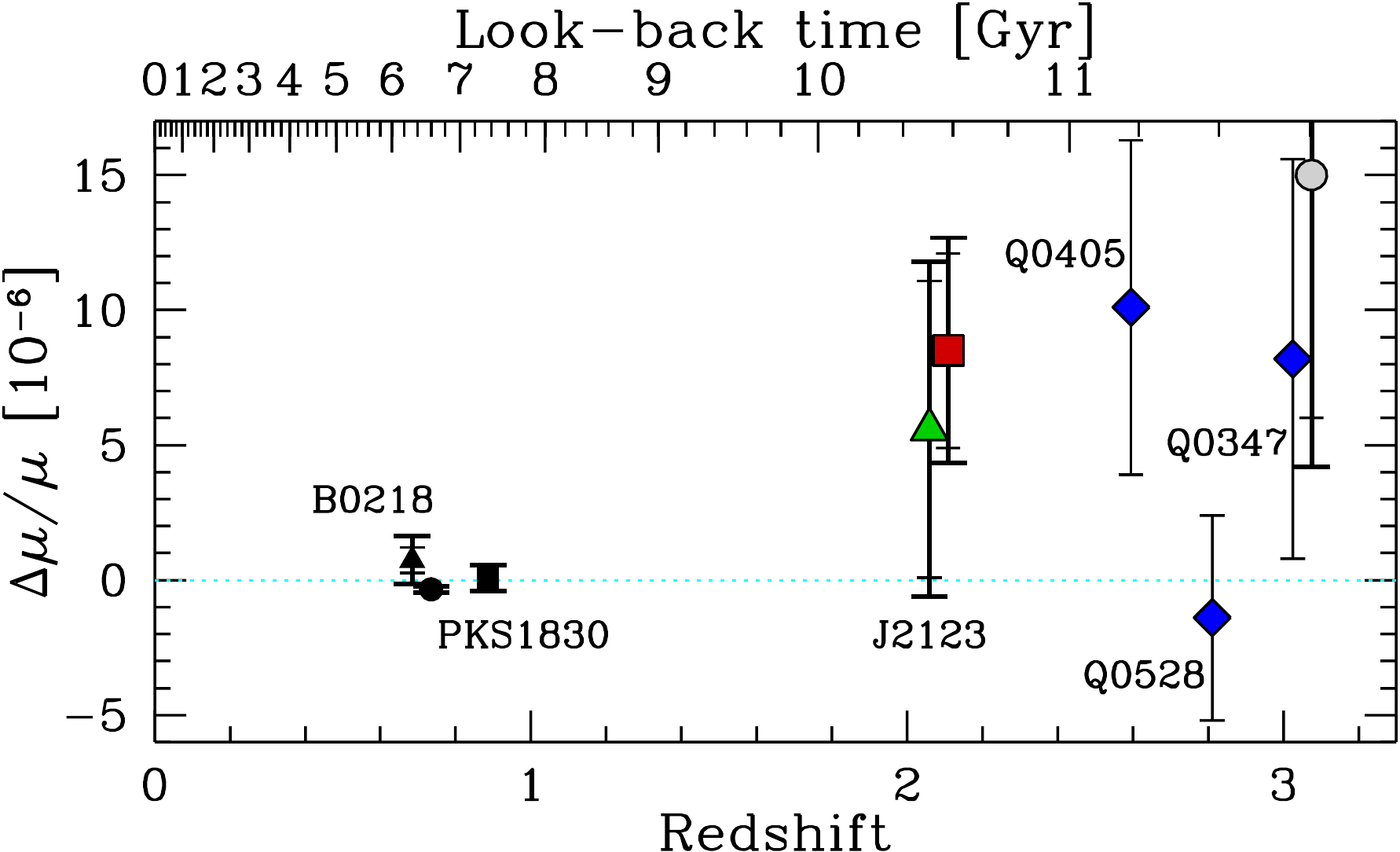}}
\caption{(Color online) Current constraints on $\Delta\mu/\mu$ on a cosmological time scale. Black points for $z<1$ refer to NH$_3$ values from objects B0218~\cite{Murphy2008,Kanekar2011} and PKS1830~\cite{Henkel2009}; blue diamonds refer to Ref.~\cite{King2008}; green triangle to Ref.~\cite{Malec2010}, \emph{i.e.} the HIRES-Keck analysis of J2123. The grey circle refers to a recent re-analysis of Q0347~\cite{Wendt2011}. The red square is the present result of our UVES-VLT analysis of J2123. All uncertainties are at $1\sigma$. Independent constraints on the same object have been offset in redshift for clarity.}
\label{Fig4}
\end{figure}

Spectral observations of highly redshifted H$_2$ absorbers are limited. Of the many thousands of known quasar systems, of which some 1000 have been classified as damped-Lyman-$\alpha$ (DLA) systems, in only some twenty is molecular hydrogen detected, but not under conditions allowing for a detailed analysis of $\mu$-variation at a competitive level. So far only high quality spectra from H$_2$ absorbing systems toward Q0347$-$383 at $z_{\mathrm{abs}}=3.02$ and Q0405$-$443 at $z_{\mathrm{abs}}=2.59$~\cite{Reinhold2006,King2008} and Q0528$-$250 at $z_{\mathrm{abs}}=2.81$~\cite{King2008} have been observed with the Very Large Telescope to yield a tight constraint on $\Delta\mu/\mu$. Recently a VLT-reinvestigation of Q0347$-$383 was also reported~\cite{Wendt2011}. In contrast, for detecting drifts in $\alpha$, spectra of atomic species (Mg~{\sc i}, Mg~{\sc ii}, Fe~{\sc ii}, Zn~{\sc ii}, Cr~{\sc ii}, Si~{\sc ii}, etc.) in almost $300$ absorption systems have been analyzed~\cite{Murphy2003}. For H$_2$, a spectrum of J1337+315 at $z_{\mathrm{abs}}=3.17$ was analyzed~\cite{Srianand2010}, but due to the low column density ($\log N(\mathrm{H}_2) = 14.1$) a constraint on $\Delta\mu/\mu$ only at the non-competitive $10^{-4}$ level was found.

In Fig.~\ref{Fig4} the now existing information on high redshift absorbing H$_2$ systems is collected; the tight constraints from the ammonia method~\cite{Murphy2008,Kanekar2011,Henkel2009} for $z<1$ are included as well. The status result of a possible variation of the proton-to-electron mass ratio can be expressed as: there exists no firm evidence for a drifting $\mu$ on a cosmological time scale, which is constrained by $\Delta\mu/\mu < 1 \times 10^{-5}$ at redshifts in the range $z=2-3$. Quantitatively, averaging the six results for the four high-redshift H$_2$ absorption systems yields $\Delta\mu/\mu = (5.2 \pm 2.2)\times 10^{-6}$, which is a slight indication (at 2.3 $\sigma$ level) for a larger $\mu$ at high redshift, but further systems should be analyzed to warrant such a claim. As for the proportionality relation of Eq.~(\ref{Eq-R}), combined with the finding of $\Delta\alpha/\alpha = (-5.4 \pm 1.2)\times 10^{-6}$~\cite{Murphy2003}, some significant drift on $\mu$ exceeding the present constraint would be expected. The result may also be interpreted as the specific assumptions of Grand Unification underlying Eq.~(\ref{Eq-R}) to be unjustified. Indeed, while the majority of the GUT scenarios predict $|R_{\mu \alpha}|$ to be large, some special cases are reported yielding a smaller proportionality constant, \emph{i.e.} $|R_{\mu \alpha}| \leq 5$~\cite{Dent2008}. The present results constraining $\Delta\mu/\mu$ may thus be used to discriminate between various possible GUT scenarios.

This work is based on observations carried out at the European Southern
Observatory (ESO) under program ID 81.A-0242 (PI Ubachs), with the
UVES spectrograph installed at the Kueyen UT2 on Cerro Paranal, Chile.
The authors acknowledge fruitful discussions with J. King and J.K. Webb (Sydney).
MTM thanks the Australian Research Council, and WU thanks the Netherlands Foundation for Fundamental Research on Matter (FOM) for research support.

\end{document}